\documentclass[prb,aps,epsf,twocolumn,showpacs,10pt,superscriptaddress]{revtex4-1}
\usepackage{graphicx,amsfonts,times,bm,amsmath,verbatim,color,array}
\usepackage{graphicx}
\usepackage{color}
\usepackage{amsthm}
\usepackage{amsmath}
\usepackage{amssymb}
\usepackage{setspace}
\usepackage{float}

\renewcommand{\[}{\left[}
\renewcommand{\]}{\right]}

\usepackage{subfigure}
\usepackage{float}
\usepackage{epsfig}
\usepackage{psfrag}
\usepackage{hyperref}
\hypersetup{
	colorlinks=true,final=true,
	linkcolor=blue,
	citecolor=blue,
	filecolor=blue,
	urlcolor=blue,
}

\begin{document}
\title{Transverse thermopower in Dirac and Weyl semimetals}

\author{Girish Sharma}
\affiliation{Department of Physics, National University of Singapore, Singapore 117551, Singapore}
\affiliation{Centre for Advanced 2D Materials and Graphene Research Centre, National University of Singapore, Singapore 117546, Singapore}
\affiliation{School of Basic Sciences, Indian Institute of Technology Mandi, Mandi-175005 (H.P.) India}
\author{Sumanta Tewari}
\affiliation{Department of Physics and Astronomy, Clemson University, Clemson, South Carolina 29634, USA}

\begin{abstract}
	Dirac semimetals (DSM) and Weyl semimetal (WSM) fall under the generic class of three-dimensional solids, which follow relativistic energy-momentum relation $\epsilon_\mathbf{k}= \hbar v_F |\mathbf{k}|$ at low energies. Such a linear dispersion when regularized on a lattice can lead to remarkable properties such as the anomalous Hall effect, presence of Fermi surface arcs, positive longitudinal magnetoconductance, and dynamic chiral magnetic effect. The last two properties arise due to the manifestation of chiral anomaly in these semimetals, which refers to the non-conservation of chiral charge in the presence of electromagnetic gauge fields. Here, we propose the planar Nernst effect, or transverse thermopower, as another consequence of chiral anomaly, which should occur in both Dirac and Weyl semimetals. We analytically calculate the planar Nernst coefficient for DSMs (type-I and type-II) and also WSMs (type-I and type-II), using a quasi-classical Boltzmann formalism. The planar Nernst effect manifests in a configuration when the applied temperature gradient, magnetic field, and the measured voltage are all co-planar, and is of distinct origin when compared to the anomalous and conventional Nernst effects. 
	Our findings, specifically a 3D map of the planar Nernst coefficient in type-I Dirac semimetals (Na$_3$Bi, Cd$_3$As$_2$ etc) and type-II DSM (PdTe$_2$, VAI3 etc), can be verified experimentally by an in-situ 3D double-axis rotation extracting the full $4\pi$ solid angular dependence of the Nernst coefficient.
\end{abstract}

\maketitle
\section{Introduction}
The well known non-crossing theorem~\cite{Neumann:1929} states that Bloch bands with the same symmetry cannot be degenerate at a generic point in the Brillouin zone, which gives rise to an avoided level crossing. However, the non-crossing theorem does not apply to bands
exhibiting non-trivial topology, which can form topologically protected band degeneracies~\cite{Chiu:2016,Volovik,YangHang:2017,ArmitageMeleVishwanath:2017}. Dirac semimetals (DSMs) and Weyl semimetals (WSMs) are celebrated examples of three-dimensional systems with topologically protected level crossing near the Fermi level, exhibiting low energy excitations with relativistic energy-momentum
relations resembling massless Dirac fermions~\cite{Murakami1:2007,Murakami2:2007,Wan:2011,Burkov1:2011,Burkov:2011, Xu:2011,Yang:2011}.
In a WSM the level crossings of non-degenerate pairs of bands can act as source and sink of Abelian Berry curvature~\cite{Xiao:2010}, and are topologically protected by a non-zero flux of Berry curvature across the Fermi surface. Nielsen and Ninomiya~\cite{Nielsen:1981, Nielsen:1983} showed that on a lattice, only an even number of Weyl points can occur, which carry opposite monopole charges such that the net monopole charge summed over all the Weyl points in the Brillouin zone vanishes. 
In a DSM, time reversal (TR) and space inversion (SI) symmetries are simultaneously preserved, and the bulk energy bands are Kramers degenerate. This
ensures that an accidental crossing between valence and conduction bands engenders a four-fold degenerate Dirac node, which can be stable in the presence of additional symmetries, such as uniaxial discrete crystal rotation symmetries $C_n$~\cite{YangNagaosa:2014, YoungZaheer:2012, SteinbergYoung:2014}.
Also the simultaneous presence of TR and SI  symmetry ensures that
the monopole charge vanishes at each crossing point. This can be contrasted with a WSM, which breaks either SI or TR symmetry. A WSM phase can be generated from a DSM by breaking either of these two symmetries, for example by applying a magnetic field, which breaks TR symmetry.

Recently there has been a surge of interest in DSMs and WSMs~\cite{Su:2015, Huang:2015, Lv:2015,Liu:2014, SyXu:2013, Neupane:2014, ZKLiu:2014, Borisenko:2014, Jeon:2014, TLiang:2014, Xiong1:2015, Xiong:2015,Wang:2012, Wang:2013,Fu:2007, Teo:2008, Guo:2011,Kim:2013,Llu:2015}, as they evince many topological transport and optical properties not shared by other 3D materials.
TR broken WSMs exhibit anomalous Hall and Nernst effects~\cite{BurkovPRL2014, Sharma:2016, Sharma:2017, TLiang:2017, Watzman:2018,Lundgren:2014, Ferrerios2017,Chernodub2018, Rana2018, Caglieris:2018,GorbarMiransky2017,DasAgarwal}, while dynamic chiral magnetic effect can be related to optical gyrotropy and natural optical activity in inversion broken WSMs~\cite{Goswami:2015,Zhong:2015}. 
More interestingly, the surface of WSMs hosts distinct Fermi arcs, and the bulk transport is characterized by negative longitudinal magnetoresistance in the presence of parallel electric
and magnetic fields due to  chiral anomaly~\cite{Goswami:2013,Bell:1969,Aji:2012,Adler:1969,Zyuzin:2012, LiangPRX:2018, Sharma2:2017, Y-YLv:2017,XHuang:2015, Shekhar:2015,BurkovPRB:2015, Zyuzin:2017, He:2014,CLZhang:2016,QLi:2016,Xiong,Hirsch,Son:2013,Kim:2014, Barnes:2016}. Very recently, the current authors along with others, proposed the planar Hall effect (PHE)~\cite{Nandi,Burkov2017} as another striking consequence of chiral anomaly in WSMs, where the effect manifests itself when the applied current, magnetic field, and the induced transverse Hall voltage all lie in the same plane, precisely in a configuration in which the conventional Hall effect vanishes. This resulted in a series of experiments, where this effect was confirmed by several groups~\cite{pheexpt1,pheexpt2,pheexpt3,pheexpt4,pheexpt5,pheexpt6,pheexpt7,pheexpt8,pheexpt9,pheexpt10,pheexpt11,pheexpt12} in Weyl and Dirac semimetals.

Even though the anomalous Nernst effect should vanish in a continuum model of Weyl fermions~\cite{Lundgren:2014}, it was shown to be both non-vanishing and measurable in both Dirac and Weyl semimetals~\cite{Sharma:2016,Sharma:2017}. In fact, a large Nernst signal has been experimentally measured in both Dirac~\cite{TLiang:2017} and Weyl semimetals~\cite{Watzman:2018, Rana2018, Caglieris:2018}, which primarily arises due to the giant Berry curvature of the Bloch bands and the Dirac dispersion, respectively. The predicted Nernst effect strictly falls into the category of anomalous or conventional Nernst response. An anomalous (conventional) Nernst effect requires the presence of Berry curvature (magnetic field) in a direction perpendicular to the plane of temperature gradient and the induced voltage. Here, we propose another type of Nernst effect, namely the planar Nernst effect (PNE), which strictly arises as a consequence of chiral anomaly, and displays properties distinct from both the conventional and anomalous Nernst effects. The planar Nernst effect can be also viewed as transverse thermopower, analogous to the conventional thermopower (Seebeck coefficient) where a thermal gradient induces a thermoelectric voltage. In the current scenario, the voltage is induced transverse to the thermal gradient. 
The effect is shown to manifest in a configuration when the applied temperature gradient, magnetic field, and the induced voltage are co-planar. The planar Nernst effect is known to occur in ferromagnetic systems~\cite{Schmid2013,Avery2012,YPu:2006}, however to the best of our knowledge has not been explored in WSMs/DSMs.
We develop a quasiclassical theory of the planar Nernst effect in Weyl and Dirac semimetals, where the  Fermi surfaces enclose nonzero fluxes of the Berry curvature in momentum space. Our findings, specifically a 3D map of the planar Nernst coefficient in type-I (Na$_3$Bi, Cd$_3$As$_2$ etc) and type-II Dirac semimetals (VAI3, PdTe$_2$ etc), can be verified experimentally by an in-situ 3D double-axis rotation extracting the full $4\pi$ solid angular dependence~\cite{TLiang2018}.
\begin{figure}
	\includegraphics[scale=0.6]{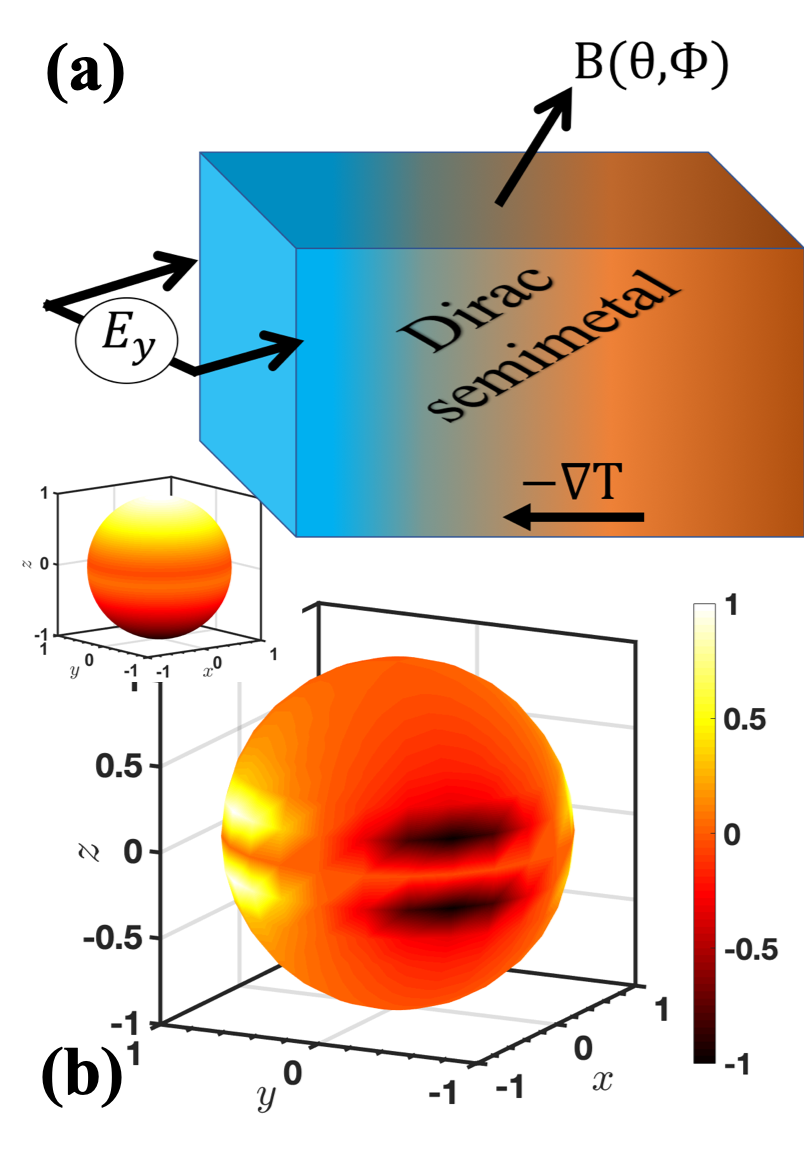}
	\caption{\textit{(a)} Schematic illustration for the measurement of the planar Nernst coefficient in Dirac semimetals. A longitudinal temperature gradient $dT/dx$ produces a chiral anomaly induced transverse electric field $E_y$ due to the co-planar component of the $\mathbf{B}$ field. As the magnetic field 
		$(\mathbf{B})$ is rotated along $\hat{\theta}$ (polar angle) and $\hat{\phi}$ (azimuthal angle) directions,  one can map the planar Nernst coefficient for a Dirac semimetal (note that the $B_z$ component in a DSM produces Weyl points, while the planar components of $\mathbf{B}$ produce the planar Nernst effect due to chiral anomaly). The 3D map of the planar Nernst coefficient can be verified experimentally by an in-situ 3D double-axis rotation extracting the full $4\pi$ solid angular dependence~\cite{TLiang2018}. \textit{(b)} Mapping the planar Nernst coefficient $\nu^{\text{pl}}$ in a type-I Dirac semimetal, showing the angular dependence w.r.t $\theta$ and $\phi$, where $\theta$ is the polar angle and $\phi$ is the azimuthal angle. The small plot on the left side shows the map of the anomalous Nernst coefficient for the same system.}
			\label{Fig:DSMexpt}
\end{figure}

\section{Planar Nernst Effect}
The conventional Nernst effect measures the transverse electrical response to a longitudinal thermal gradient in the presence of an out of plane magnetic field and absence of a charge current i.e. $E_y=-\nu$ $dT/dx$, where $\nu$ is defined to be the Nernst coefficient and $-dT/dx$ is the temperature gradient applied along the $x$ axis, and $\mathbf{B}=B\hat{z}$. 
In terms of the conductivity tensors $\hat{\sigma}$ and $\hat{\alpha}$, the Nernst coefficient $\nu$ can be derived to be
\begin{align}
\nu =   \frac{E_y}{(-dT/dx)} = \frac{\alpha_{xy}\sigma_{xx} - \alpha_{xx}\sigma_{xy}}{\sigma_{xx}^2 + \sigma_{xy}^2},
\label{nernst_eqn}
\end{align}
A conventional Nernst effect requires a non-zero component of the magnetic field parallel to the $\hat{\mathbf{E}} \times {\nabla T}$ plane, which provides the Lorentz force to the quasiparticles. On the other hand, an anomalous Nernst effect (due to the Berry phase) does not explicitly require a magnetic field, but rather requires a non-zero component of the Berry curvature, again parallel to the $\hat{\mathbf{E}} \times {\nabla T}$ plane. Both of these effects have been well studied and experimentally observed in WSMs and DSMs~\cite{Sharma:2016,Sharma:2017,TLiang:2017,Caglieris:2018,Watzman:2018,Rana2018}. Here, we predict a third type of Nernst response, namely the planar Nernst effect, which should also occur both in Dirac and Weyl semimetals. Unlike the conventional and the anomalous Nernst effects, the planar Nernst effect is characterized by co-planar $\mathbf{E}$, $\nabla T$, and $\mathbf{B}$ fields, and is a direct consequence of chiral anomaly in 3D Dirac materials. Fig.~\ref{Fig:DSMexpt} schematically illustrates the measurement of the planar Nernst coefficient in Dirac semimetals. A longitudinal temperature gradient $dT/dx$ produces a transverse electric field $E_y$ due to chiral anomaly as a result of the co-planar component of the $\mathbf{B}$ field. As the magnetic field $(\mathbf{B})$ is rotated along $\hat{\theta}$ and $\hat{\phi}$ directions, where $\hat{\theta}$ is the polar angle and $\hat{\phi}$ is the azimuthal angle, one can map the planar Nernst coefficient for a Dirac semimetal (note that the $B_z$ component in a DSM produces Weyl points, while the $B_x$ and $B_y$ components produce the planar Nernst effect due to chiral anomaly).
Here we will calculate the planar Nernst coefficient $\nu^{\text{pl}}$ using the quasi-classical Boltzmann formalism.  in the relaxation time approximation, accounting for contributions from an external magnetic field and Berry curvature. 

In the presence of Berry curvature $\mathbf{\Omega}_{\mathbf{k}}$, the semi-classical equations of motion for an electron are modified~\cite{Sundurum:1999,Niu:2006,Son:2012,Duval:2006}. We need to account for the Berry curvature contributions while solving the quasi-classical Boltzmann equations. The steady state Boltzmann equation in the relaxation time approximation is given by
\begin{eqnarray}
(\mathbf{\dot{r}}\cdot \nabla_{\mathbf{r}} + \mathbf{\dot{k}}\cdot \nabla_{\mathbf{k}})f_{\mathbf{k}} =-\frac{f_{\mathbf{k}}-f_{eq}}{\tau_\mathbf{k}},
\label{boltz_basic_eqn}
\end{eqnarray}
where $\tau_\mathbf{k}$ is the scattering time, $f_{eq}$ is the equilibrium Fermi-Dirac distribution function, and $f_\mathbf{k}$ is the distribution function of the system in the presence of perturbations. We point out that the scattering time $\tau_\mathbf{k}$ actually depends on the nature of underlying impurities, and can lead to a non-trivial energy dependence. Nevertheless, for simplicity, for a finite chemical potential, we treat $\tau$ to be approximately energy independent~\cite{Sharma:2016}.  { We also take an approximate momentum dependence of scattering time ($\tau \rightarrow \tau (1-\exp(-4(|k_z| - d)^2 / d^2))$) such that internode scattering dominates over intranode scattering because for longitudinal magnetoconductance the internode scattering is supposed to be the dominant scattering mechanism. 
} 
\begin{figure}
	\includegraphics[scale=.76]{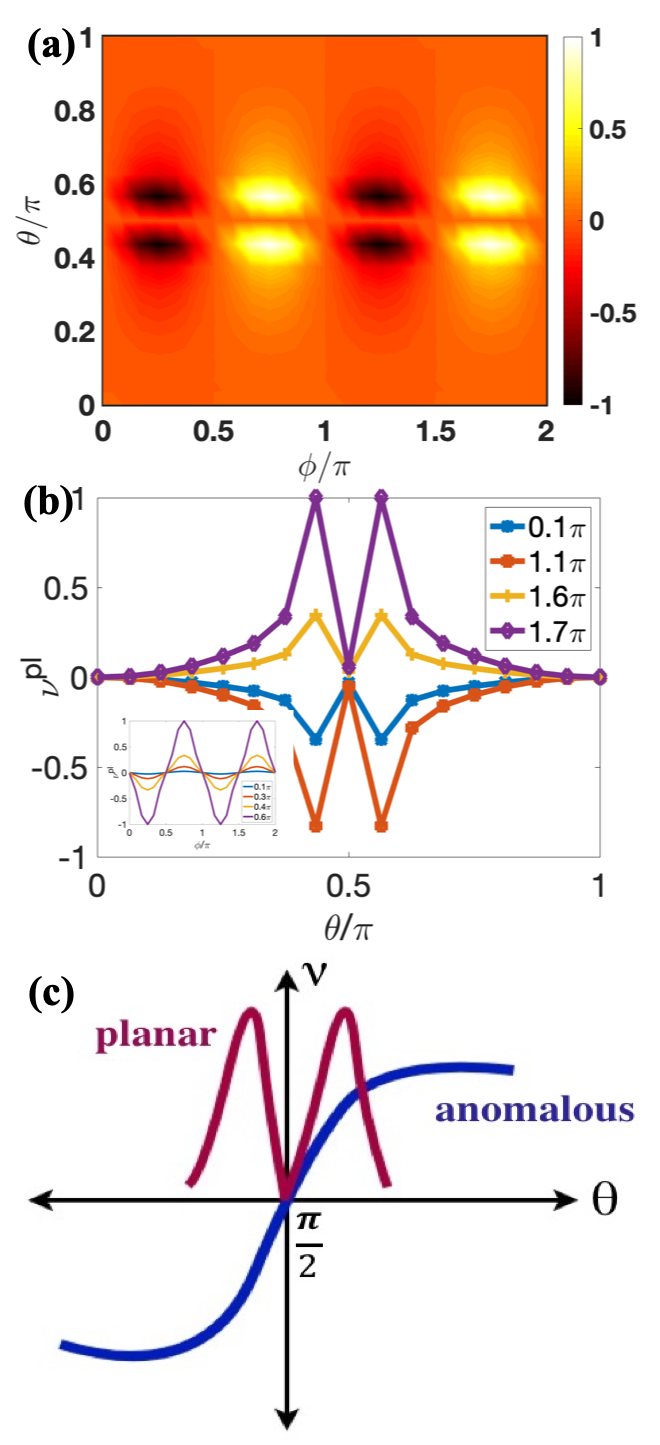}
	\caption{\textit{(a)} Density plot of the planar Nernst coefficient $\nu^{\text{pl}}$ for a type-I DSM.   \textit{(b)} Plot of $\nu^{\text{pl}}$  as a function of $\theta$ for various values of $\phi$. We observe a peak in the Nernst coefficient away from $\theta=\pi/2$. The inset shows the plot of $\nu^{\text{pl}}$  as a function of $\phi$ for various values of $\theta$, showing a $\sin\phi\cos\phi$ behavior. \textit{(c)} Sketch of qualitative $\theta$ dependence of the anomalous and planar Nernst coefficient.}
	\label{Fig:DSM:planarNernst}
\end{figure}

\section{Solution for the Boltzmann equation and planar Nernst coefficient}
In the presence of Berry curvature $\mathbf{\Omega}_{\mathbf{k}}$, the semi-classical equation of motion for an electron takes the following form~\cite{Niu:2006, Sundurum:1999}
\begin{eqnarray}
\mathbf{\dot{r}} = \frac{1}{\hbar} \frac{\partial\epsilon(\mathbf{k})}{\partial\mathbf{k}} + {\mathbf{\dot{k}}}\times\mathbf{\Omega}_{\mathbf{k}},
\label{r_dot_eqn_1}
\end{eqnarray}
where $\mathbf{k}$ is the crystal momentum, $\epsilon(\mathbf{k})$ is the energy dispersion. The first term in Eq.~\ref{r_dot_eqn_1} is the familiar relation between semi-classical velocity $\mathbf{\dot{r}}$ and the band energy dispersion $\epsilon(\mathbf{k})$. The second term is the anomalous transverse velocity term originating from $\mathbf{\Omega(\mathbf{k})}$. In the presence of electric and magnetic fields we have the standard relation: $\mathbf{\dot{p}} = e\mathbf{E} + e\mathbf{\dot{r}}\times\mathbf{B}$. These two coupled equations for $\mathbf{\dot{r}}$ and $\mathbf{\dot{p}}$ can be solved together to obtain~\cite{Duval:2006, Son:2012}
\begin{align}
&\mathbf{\dot{r}} = D(\mathbf{B},\Omega_{\mathbf{k}}) \[\mathbf{v}_{\mathbf{k}} + \frac{e}{\hbar} (\mathbf{E}\times\Omega_{\mathbf{k}}) + \frac{e}{\hbar}(\mathbf{v}_{\mathbf{k}}\cdot\Omega_{\mathbf{k}})\mathbf{B}\] \label{rdot_eqn}\\
&\hbar\mathbf{\dot{k}} = D(\mathbf{B},\Omega_{\mathbf{k}}) \[e\mathbf{E} + \frac{e}{\hbar} (\mathbf{v}_{\mathbf{k}}\times B) + \frac{e^2}{\hbar}(\mathbf{E}\cdot\mathbf{B})\mathbf{\Omega_{\mathbf{k}}}\], \label{pdot_eqn}
\end{align}
where $D(\mathbf{B},\Omega_{\mathbf{k}}) = (1+e (\mathbf{B}\cdot\Omega_{\mathbf{k}})/\hbar)^{-1}$.  We will denote $ D(\mathbf{B},\Omega(\mathbf{k})) \equiv \mathcal{D}$ without explicitly pointing out the implied $\mathbf{B}$ and $\Omega(\mathbf{k})$ dependence. In Eq.~\ref{rdot_eqn} and Eq.~\ref{pdot_eqn}, we have also defined $\mathbf{v}_{\mathbf{k}}=\hbar^{-1}\partial\epsilon_{\mathbf{k}}/\partial\mathbf{k}$ to be the band-velocity. 

Here, we are interested in the configuration $\nabla T = \frac{d T}{dx} \hat{x}$, $\mathbf{E}=0$, but the magnetic field $\mathbf{B} = (B \cos \phi, B \sin \phi, 0)$ is along the in-plane direction. The Boltzmann equation takes the following form 
\begin{align}
&\left[v_x + \frac{e}{\hbar} (\mathbf{v}\cdot \Omega) B \cos\phi\right] \left(\frac{\epsilon - \mu}{T}\right) \left(\frac{dT}{dx}\right) \left(-\frac{\partial f_{eq}}{\partial \epsilon}\right) + \nonumber \\ 
& \frac{e B}{\hbar^2} \left[-v_z\sin\phi\partial_{k_x} + v_z\cos\phi\partial_{k_y} + (v_x\sin\phi-v_y\cos\phi)\partial_{k_z}\right]f_\mathbf{k}\nonumber\\
& = -\frac{f_\mathbf{k}-f_{eq}}{\mathcal{D} \tau}.
\end{align}
We solve the above Boltzmann equation, using the following ansatz~\cite{Sharma:2016,Lundgren:2014}
\begin{align}
&f_\mathbf{k} -f_{eq} = \nonumber\\  
&- \left[\left( v_x +\frac{e B \cos\phi}{\hbar} (\mathbf{v}\cdot\Omega)\right)\mathcal{D}\tau\left(\frac{\epsilon - \mu}{T}\right) \frac{dT}{dx}\right] \left(-\frac{\partial f_{eq}}{\partial \epsilon}\right)\nonumber\\ 
&+ \mathbf{v}\cdot\Lambda \left(-\frac{\partial f_{eq}}{\partial \epsilon}\right). 
\end{align}
We solve explicitly for the correction factor $\Lambda$, but we examine that this correction factor is orders of magnitude smaller than the other terms (in the limit when the Boltzmann equation is valid i.e. $\mu\gg k_BT, \hbar\omega_c$), and we thus retain only leading order terms in the distribution function $f_\mathbf{k}$. 
We can then write the following relation for the charge current 
\begin{align}
\mathbf{J}  = -e \int{\frac{d^3\mathbf{k}}{(2\pi)^3} \mathcal{D}^{-1} \dot{\mathbf{r}}f_{\mathbf{k}}} + \frac{k_B e \nabla T}{\hbar} \times \int{\frac{d^3\mathbf{k}}{(2\pi)^3}\Omega_{\mathbf{k}} s_\mathbf{k}}.
\label{Eq:J1}
\end{align}
The quantity $s_{\mathbf{k}} = -f_{eq}\log f_{eq} - (1-f_{eq} \log(1-f_{eq}))$ is the entropy density for the Weyl/Dirac electron gas~\cite{ZhangTewari:2008}. The second term in the above equation describes a purely anomalous Nernst response in the absence of any magnetic field.  The first term in the above equation describes the Nernst response in the presence of the magnetic field. This is the quantity which is of interest to us here.  We can then read the planar Peltier coefficient $\alpha_{xy}^{\text{pl}}$ as 
\begin{align}
\alpha_{xy}^{\text{pl}} = -e &\int{\frac{d^3\mathbf{k}}{(2\pi)^3} \left(\frac{e (\mathbf{v}\cdot \Omega) B \sin \phi}{\hbar}\right) b_0 (v_x + \frac{eB}{\hbar} \cos\phi (\mathbf{v}\cdot\Omega))}\nonumber\\
& {\left(-\frac{\partial f_{eq}}{\partial \epsilon}\right) },
\label{Eq:alphaxyplanar1}
\end{align}
where $b_0 = \mathcal{D} \tau \left(\frac{\epsilon - \mu}{T}\right)$. The Nernst coefficient can be evaluated from Eq. 2 of the main text, where $\alpha_{xx}$, $\sigma_{xx}$ and $\sigma_{xy}$ can be evaluated in a similar fashion for the same experimental configuration. When the Weyl cones are not tilted, the expression simplifies to 
\begin{align}
\alpha_{xy}^{\text{pl}} = -e &\int{\frac{d^3\mathbf{k}}{(2\pi)^3} \left(\frac{e^2 (\mathbf{v}\cdot \Omega)^2 B^2 \sin \phi\cos\phi}{\hbar^2}\right) b_0 }
{\left(-\frac{\partial f_{eq}}{\partial \epsilon}\right) }.
\label{Eq:alphaxyplanar2}
\end{align}

\section{Planar Nernst effect in Dirac semimetals}
We will now discuss the planar Nernst effect in Dirac semimetals. 
We begin with the effective low energy Hamiltonian for a type-I Dirac semimetal Cd$_3$As$_2$, in the basis $|s, \uparrow\rangle$, $|p_x+ip_y, \uparrow\rangle$, $|s,\downarrow\rangle$, $|p_x-ip_y,\downarrow\rangle$, which can be written as~\cite{YangNagaosa:2014,Hashimoto:2016, Cano:2016}
\begin{eqnarray}
H_{\mathbf{k}} &= a(\mathbf{k})\sigma_zs_0 + b(\mathbf{k})\sigma_xs_z  + c(\mathbf{k})\sigma_ys_0.
\label{Eqn_H_k_DSM}
\end{eqnarray}
In Eq.~\ref{Eqn_H_k_DSM}, $\sigma$ and $s$ are Pauli matrices representing the orbital degree of freedom and spin degree of freedom respectively. The matrix $s_0\equiv I_2$ is the two-dimensional identity matrix in spin space. The functions $a(\mathbf{k})-c(\mathbf{k})$ are defined as
\begin{align}
a(\mathbf{k})&=m_0-m_1 k_z^2-m_{2}(k_x^2+k_y^2),\label{Eq_a_DSM}\\
b(\mathbf{k})&=\eta k_x,\label{Eq_b_DSM}\\
c(\mathbf{k})&=-\eta k_y,\label{Eq_c_DSM}.
\end{align}
Here we have only included terms up to the order $\mathcal{O}(k^2)$.
The parameters $m_0$, $m_1$, $m_2$, $\eta$ depend on the material. Specifically for Cd$_3$As$_2$ ab-inito calculations upto order $k^2$ yield $m_0=.02eV$, $m_1=18.77eV \AA^2$, $m_2=13.5 eV \AA^2$, $\eta=0.89 eV\AA$~\cite{Cano:2016}.  This Hamiltonian produces two Dirac points at $\mathbf{K}=(0,0,\pm\sqrt{m_0/m_1})$ where the energy dispersion exactly vanishes. The effect of an external magnetic field $\mathbf{B}$, coupling to the spin degree of freedom can be now introduced by adding the Zeeman term~\cite{Hashimoto:2016} $H_Z = -\mu (b_x (\sigma_{0}+\sigma_z)s_x + b_y(\sigma_{0}+\sigma_z)s_y+b_z \sigma_0s_z)$ in the Hamiltonian, where $\mathbf{B}=(b_x,b_y,b_z)$, and $\mu$ is the magnetic moment. With the applied magnetic field the Hamiltonian now produces a TR broken Weyl semimetal. {The Bloch electrons also carry an orbital magnetic moment, given by}
\begin{align}
m_n^{\gamma} = \frac{i e}{2\hbar}\sum_{n'\neq n}\frac{\langle n|\partial H / \partial k_\alpha|n'\rangle \langle n'|\partial H/\partial k_\beta | n\rangle}{E_n - E_{n'}}
\end{align} 
{This also gives rise to a Zeeman like contribution and the energy spectrum is shifted as $E_n \rightarrow E_n - m_n^{\gamma}b^\gamma$.}

{ It is also important to understand how do the Weyl points evolve in this model as a function of the magnetic field. When $\theta =0$, the Weyl points are separated along the $k_z$ direction and are loacted at $(0,0,\pm\sqrt{m_0\pm b_z/m_1})$ and occur at the same energy. When $\theta\neq 0$, the position of the Weyl nodes remains unchanged and they only move along the energy axis in a manner such that inversion symmetry is preserved.}
It is important that $b_z\neq 0$ in order to generate Weyl points from the Dirac nodes.  In Eq.~\ref{Eq:alphaxyplanar1} we calculated the planar Nernst coefficient for an in-plane magnetic field $(b_x,b_y,0)$, however it is essential that in order to observe a planar Nernst effect in a Dirac semimetal, we need a finite magnetic field along the $z$ direction. This is because, we need a finite flux of Berry curvature, which is generated by a magnetic field along the $z$ axis due to the generation of Weyl points. We therefore have the configuration $\mathbf{B}=B(\sin\theta\cos\phi, \sin\theta\sin\phi, \cos\theta)$, $\nabla T = (-dT/dx,0,0)$. Following the standard procedure as mentioned earlier, we can straightforwardly extend the solution to the Boltzmann equation.

Fig.~\ref{Fig:DSMexpt} shows the 3D mapping of the planar Nernst coefficient for a type-I Dirac semimetal, as a function of polar and azimuthal angle, for a constant $|\mathbf{B}|$, which is rotated in space. Such a 3D map of the planar Nernst coefficient can be verified experimentally by an in-situ 3D double-axis rotation extracting the full $4\pi$ solid angular dependence~\cite{TLiang2018}. Fig.~\ref{Fig:DSM:planarNernst} shows the density plot of the planar Nernst coefficient, and its specific dependence on the polar and azimuthal angles. 
The planar Nernst coefficient $\nu^{\text{pl}}$ at a constant $\theta$ exhibits the $\sin\phi\cos\phi$ behaviour (when $\theta\neq\{0,\pi/2,\pi\}$). Strikingly, we find that $\nu^{\text{pl}}$ shows a peculiar $\theta$ dependence at a constant $\phi$ (when $\phi\neq \{0,\pi/2,\pi,3\pi/2\}$). We see that slightly away from $\theta=\pi/2$, the planar Nernst coefficient exhibits a peak, which does not change sign as one moves across $\theta=\pi/2$ (i.e. negative $B_z$ to positive $B_z$). This can be understood as follows: a non-zero $B_z$ field  (i.e. around $\theta=\pi/2$) produces Weyl points and thereby a sharp peak in the Berry flux.  
Holding the magnitude of the applied magnetic field constant, a small $B_z$ implies that $B_x$ and $B_y$ are large enough, and the combination of the Berry curvature and the in-plane magnetic field gives rise to a large planar Nernst signal. Because $\nu^{\text{pl}}$ depends on $|\Omega|$, which is the same for a finite $\pm B_z$, the signal does not change sign as one goes from positive $B_z$ to negative $B_z$. 
Although the anomalous Nernst coefficient is also known to show a step like behaviour near $B_z=0$, it is accompanied with a sign change~\cite{Sharma:2017,TLiang:2017}. The conventional Nernst coefficient (even though it is small due to Sondheimer's cancellation~\cite{Sharma:2017}), also shows a similar sign change as one moves across $B_z=0$. 
This is a very important distinguishing feature of the planar Nernst coefficient (see Fig.~\ref{Fig:DSM:planarNernst}). 

Further, the planar Nernst coefficient does not satisfy the antisymmetric property $\alpha_{xy}=-\alpha_{yx}$. This is because the origin of the planar Nernst effect is linked to chiral anomaly, unlike usual Lorentz force term.
Thus, the planar contribution and its peculiar angle dependence can be easily extracted by rotating the field along $\phi$ and $\theta$, and subtracting the other two contributions.
\begin{figure}
	\includegraphics[scale=.67]{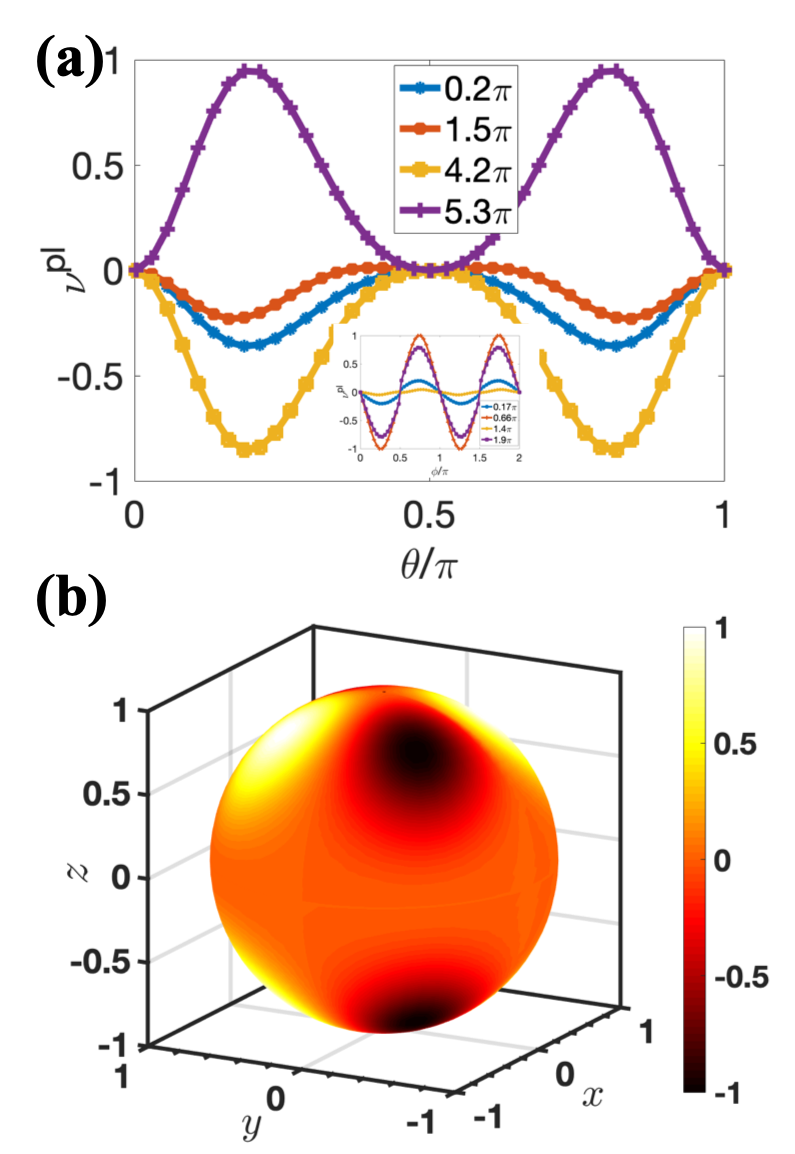}
	\caption{\textit{(a)} Map of the full angular dependence of the planar Nernst coefficient $\nu^{\text{pl}}$  for a type-II DSM. \textit{(b)} The planar Nernst coefficient $\nu^{\text{pl}}$ as a function of $\theta$ at various azimuthal angles, showing a double peaked behavior around $\theta=\pi/2$. The qualitative similarity with type-I DSM can be observed. The inset shows the $\sin\phi\cos\phi$ behavior w.r.t the angle $\phi$.}
		\label{DSMlineartype2}
\end{figure}

Having discussed the planar Nernst effect in the type-I DSM, we will also briefly discuss this effect in the type-II DSM. 
Type-II DSMs are characterized by the occurrence of Dirac nodes at TR invariant momenta points in the Brillouin zone. The linearized low-energy Hamiltonian is given by 
\begin{align}
H = (-k_x \tau_z\sigma_y + k_y\tau_z\sigma_x + k_z\tau_y\sigma_{0}),
\end{align}
where $\tau$ and $\sigma$ are the orbital and spin degrees of freedom respectively. The effect of the Zeeman field is given by $H_Z = \sum_{i=1}^{3} b_i\sigma_i$, where $b_i$'s are the components of external magnetic field i.e. $\mathbf{b}=b(\sin\theta\cos\phi,\sin\theta\sin\phi,\cos\phi)$. The energy dispersion of the lowest bands, which touch at the Weyl points, is given by 
\begin{widetext}
	\begin{align}
	E_\pm = \pm\sqrt{(b^2 + k^2) - \sqrt{b^2(k_x^2 + k_y^2 + 4k_z^2 + \cos(2\phi) (k_y^2 - k_x^2) - 2\cos(2\theta) (k_y\cos\phi - k_x \sin\phi)^2 - 2k_xk_y \sin (2\phi))}},
	\label{Eqn_theta_piby2}
	\end{align}
\end{widetext}
where $b^2 = \sum_{i=1}^3 b_i^2$. Since the above equation is quite complicated, we will carefully examine the position and evolution of Weyl points below. We will therefore examine a few cases in order to track the evolution of Weyl points. The band touching condition is generically given by $E_+=E_-$.

(i) When $\phi=0$, i.e. the magnetic field is applied along the $x-z$ plane, we have Weyl nodes at $(0,0,\pm b)$ for $\theta\neq \pi/2$, and when $\theta=\pi/2$, we have a nodal line in the $k_x=0$ plane~\cite{BurkovPRL2018}. The band touching condition when $\phi=0$ is
\begin{align}
(b^2+k^2)^2 = b^2 (2k_y^2(1- \cos 2\theta)+4k_z^2),
\label{Eq_phi_zero_case}
\end{align}

(ii) When \textit{$\theta = \pi/2$} the band touching condition is 
\begin{align}
(b^2+k^2)^2 = 4b^2 ((k_x \sin\phi - k_y \cos\phi)^2+k_z^2),
\end{align}
For $\phi=0$ or $\phi=\pi$, we obtain the following equation in the $k_x=0$ plane
\begin{align}
b^2+k_y^2+k_z^2 = 2b \sqrt{k_y^2+k_z^2},
\end{align}
which clearly defines a nodal line in the $k_x=0$ plane. We can generalize this result as follows: Consider a rotation of the $k_x-k_y$ axis by an angle $\pi/2-\phi$. We have $k_x' = k_x \sin\phi - k_y \cos\phi$, $k_y' = k_x\cos\phi +k_y\sin\phi$, and $k_z'=k_z$. Eq.~\ref{Eqn_theta_piby2} can be expressed as 
\begin{align}
(b^2+k'^2)^2 = 4b^2 (k_x'^2+k_z'^2)
\end{align}
The above equation defines a nodal line in the $k'_y=0$ plane. For example when $\phi=\pi/2$, there is a nodal line in the $k_y=0$ plane. Therefore, when $\theta=\pi/2$, we always have a nodal line irrespective of the value of $\phi$. The plane at which the nodal line occurs however depends on the value of $\phi$.

(iii) Treating the generic $\phi$ case: The band touching condition can be simplified as follows after a few straightforward algebraic manipulations to be
\begin{align}
(b^2+k^2)^2 =b^2(2(k_y\cos\phi - k_x\sin\phi)^2 (1-\cos(2\theta) ) + 4 k_z^2)
\end{align}
By a rotation of axis, as in the previous case, we can rewrite this as 
\begin{align}
(b^2+k'^2)^2 =b^2(2 k_x'^2 (1-\cos(2\theta) ) + 4 k_z'^2)
\end{align}
The above equation is identical to Eq.~\ref{Eq_phi_zero_case} (i.e. case (i) when $\phi=0$). The difference is that the $k_x-k_y$ plane has been rotated by an angle $\phi$. Since the Weyl nodes in case (i) generically occur at $k_x=0$ and $k_y=0$, the location of Weyl nodes in the current case remains invariant even for finite $\phi$ because the rotated coordinate system shares the same origin. 

We can thus conclude that: for magnetic field along any direction we have either Weyl nodes (when $\theta\neq\pi/2$), or a nodal line (when $\theta=\pi/2$). The Weyl nodes always occur at $(0,0,\pm b)$ irrespective of the value of $\phi$. The chirality of the Weyl points switch as $\theta$ changes across the the angle $\theta=\pi/2$. The nodal line occurs along the plane- $x\cos\phi + y\sin \phi=0$.
\begin{figure}
	\includegraphics[scale=.4]{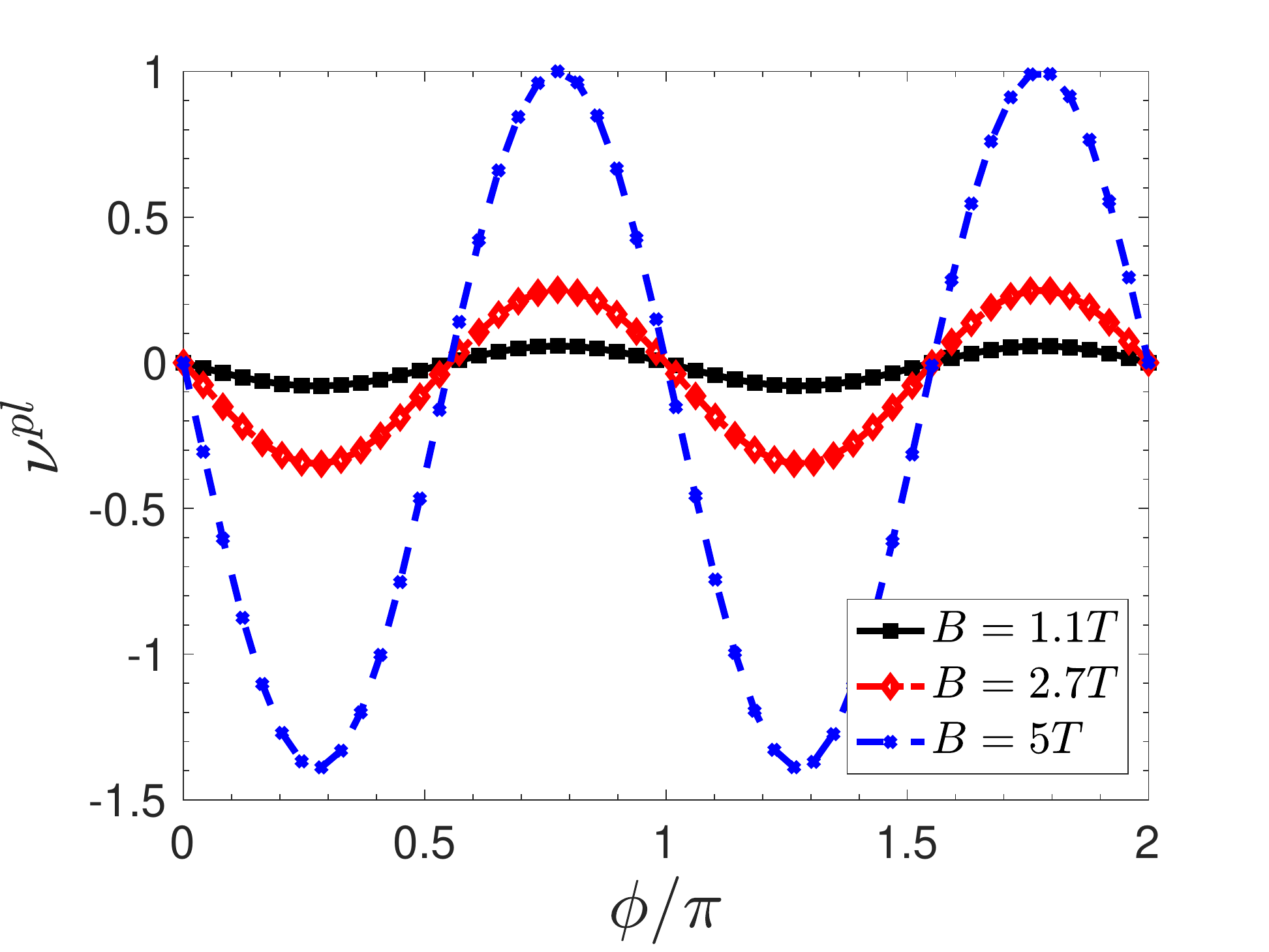}
	\includegraphics[scale=.4]{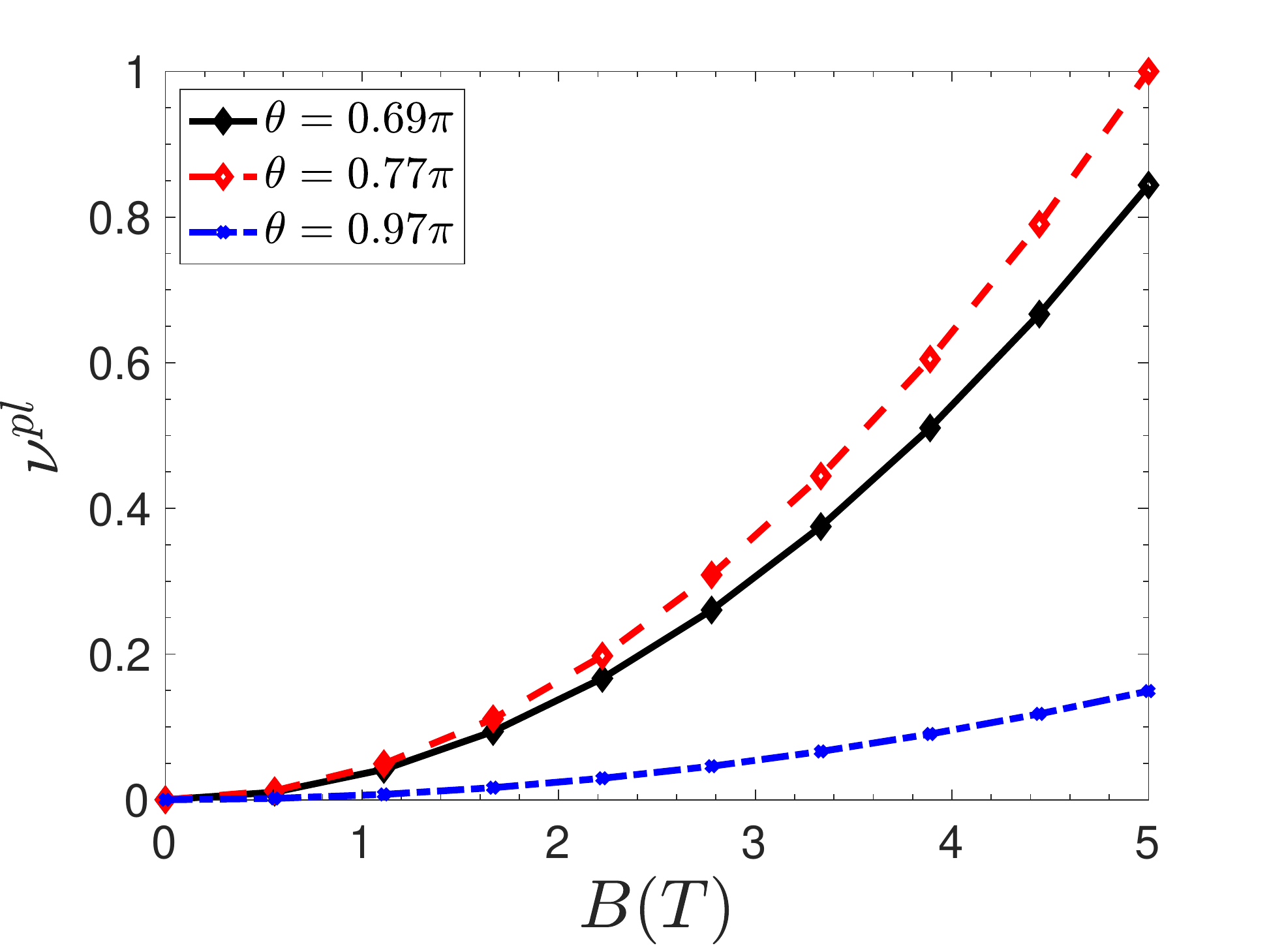}
	\caption{\textit{Top:} Planar Nernst coefficient in inversion asymmetric type-II Weyl semimetal displaying the $\cos\phi\sin\phi$ feature. \textit{Bottom:} Planar Nernst coefficient as a function of magnetic field. The linear in $B$ feature disappears here because as
		in the effects of tilts cancel out among all the nodes. We chose $t_x=t/2$, $m=2t$, $k_0=\pi/2$, $\gamma=2.5t$.}
	\label{Fig:Nernst_Weyl}
\end{figure}

Fig.~\ref{DSMlineartype2} shows  the map of the full angular dependence of the planar Nernst coefficient $\nu^{\text{pl}}$  for a type-II DSM. The planar Nernst coefficient $\nu^{\text{pl}}$ as a function of $\theta$ also shows a double peaked behaviour around $\theta=\pi/2$.
 
\section{ Planar Nernst effect in inversion asymmetric Weyl semimetals}
Here we will examine the planar Nernst effect in Weyl semimetals (inversion symmetry broken), both type-I and type-II. 
The low energy Hamiltonian for inversion symmetry breaking Weyl semimetal is given by~\cite{MccormickPRB2017} 
\begin{align}
H &= \gamma (\cos 2k_x - \cos k_0) (\cos k_z - \cos k_0) \nonumber\\
&- (m(1-\cos^2 k_z - \cos k_y) + 2t_x (\cos k_x - \cos k_0))\sigma_x \nonumber\\
&- 2t \sin k_y \sigma_y - 2t \cos k_z \sigma_z,
\end{align}
$\sigma$ being Pauli matrices int he orbital space. When $\gamma= 0$, the above equation describes a inversion asymmetric WSM with four nodes located at $k = (\pm k_0,0, \pm \pi/2)$. When the tilt parameter $\gamma> 2t$, we enter into a type-II WSM phase. The anomalous Nernst contribution from an inversion asymmetric, but TR preserving WSM is negligible, because the Berry phase effects from all the valleys cancel out.
In the present case of planar Nernst effect, we just require a strictly in-plane magnetic
field, as we already begin with Weyl points (unlike the DSM
where a non-zero $B_z$ is essential to split a Dirac point into
Weyl points). Such a configuration results in vanishing conventional Nernst effect, and thus the problem of disentangling
the planar Nernst contribution from the total Nernst effect does not arise. In Fig.~\ref{Fig:Nernst_Weyl} we plot the planar Nernst coefficient in inversion asymmetric type-II Weyl semimetal displaying the $\cos\phi\sin\phi$ feature. We also plot the planar Nernst coefficient as a function of the applied magnetic field. 
We do not get a linear in $\mathbf{B}$ dependence even for tilted cones (unlike earlier studies on the Hall/longitudinal conductivity~\cite{Sharma2:2017,Nandi}), as in the effects of tilts cancel out in the inversion asymmetric WSM model.

\section{Conclusions:}
 In this work we have presented a quasi-classical theory of chiral anomaly induced planar Nernst effect (transverse thermopower) in Dirac and Weyl semimetals.
We derived an analytical expression for the planar Nernst coefficient and also illustrated its generic behaviour for generic DSMs (type-I and type-II), and type-I and type-II WSMs. The planar Nernst effect manifests in a configuration when the applied temperature gradient, magnetic field, and the measured voltage are co-planar, and is of distinct origin when compared to the anomalous and conventional Nernst effects. We point out distinctive features of the planar Nernst coefficient in a Dirac semimetal, which exhibits a peak around $\theta = \pi/2$ as the magnetic field is rotated, with no change of sign. Such a feature can be sharply contrasted to the anomalous and conventional Nernst effect, which shows a sign change at $\theta=\pi/2$. 
Our findings, specifically a 3D map of the planar Nernst coefficient, can be verified experimentally by an in-situ 3D double-axis rotation extracting the full $4\pi$ solid angular dependence~\cite{TLiang2018}. { Experimentally it is known that jetting effect can complicate the measurement of magnetoresistance due to extrinsic reasons even in the absence of intrinsic mechanism such as chiral anomaly. In principle this may complicate measurement of planar Hall or Nernst effects as well. However in recent work~\cite{Ong2018}, it has been shown that by careful measurement of voltage drops along the mid-ridge and edges of the sample, and also by numerical simulation, one can eliminate extrinsic jetting distortions.}

\textit{Note added:} During the final stages of the preparation of this manuscript, we came across the preprint~\cite{Nag2018}, which discusses the planar Nernst coefficient, but only for multi-Weyl semimetals. 

ST acknowledges support from ARO Grant No: (W911NF16-1-0182). GS acknowledges CA2DM-NUS computing resources. 

\end{document}